\documentclass{article}

\newcommand{\bfr}{\begin{flushright}}
\newcommand{\efr}{\end{flushright}}

\makeatletter
\def\blfootnote{\xdef\@thefnmark{}\@footnotetext}
\makeatother

\usepackage{graphicx}
\usepackage{float}
\usepackage{subfigure}
\usepackage{epstopdf}
\usepackage{amsmath,amssymb}
\usepackage{verbatim}
\usepackage{multirow}
\usepackage{comment}

\usepackage{color}

\begin{document}
\title{Elastic proton-proton scattering from ISR \\
	to LHC energies, focusing on the dip region
\thanks{Presented at  the Low x workshop, May 30 - June 4 2013, Rehovot and Eilat, Israel}}

\author{
	T. Cs\"org\H{o}$^{ 1}$\blfootnote{csorgo.tamas@wigner.mta.hu}, 
	R. J. Glauber$^{ 2}$,
	 and F. Nemes$^{ 1,3}$  \\
	\smallskip\\
	{\small 
	$^{1}$Wigner Research Centre for Physics}\\
	{\small H-1525 Budapest 114, P.O.Box 49, Hungary}\\
	\smallskip\\
	{\small  $^{2}$
	Lyman Laboratory of Physics, Harvard University}\\
	{\small 17 Oxford St, Cambridge, MA02138, USA}\\
	\smallskip\\
	{\small $^{3}$ CERN, CH-1211 Geneva 23, Switzerland}
	\smallskip\\
	\null
}


\maketitle

\begin{abstract} 
The differential cross-section of elastic proton-proton
collisions is studied at ISR and LHC energies, utilizing a quark-diquark
model, that generalizes earlier models of Bialas and Bzdak, and, in addition,
a model of Glauber and Velasco.  These studies suggest that the increase of
the total pp cross-section is mainly due to an increase of the separation of
the quark and the diquark with increasing energies. Within the investigated
class of models, two simple and model-independent phenomenological relations
were found, that connect the total pp scattering cross-section $\sigma_{tot}$
to the effective quark, diquark size and their average separation, on one
hand, and to  the position of the dip of the differential cross-section, on
the other hand. The latter $t_{dip} \sigma_{tot} \simeq C$ relation can be
used to predict $t_{dip}$, the position of the dip of elastic pp scattering
for future colliding energies, and for other reactions, where $\sigma_{tot}$
is either known or can be reliably estimated.  
\end{abstract}

\vfill
{PACS: 13.75.Cs, 13.85.-t, 13.85.Lg, 13.85.Dz }
\vfill

\eject

\section{Diffraction - a historical perspective}

Diffractive scattering of electrons on various nuclei provided important
insights to the charge density distributions of spherical nuclei. The detailed
analysis resulted in simple observations by Hofstadter and colleagues, that
were summarized in the Nobel lecture of Hofstadter as follows:

\begin{itemize}
\item The volume of spherical nuclei is proportional to the mass number $A$.
\item The surface thickness is constant, independent of $A$.
\end{itemize}

These observations revealed structures in atomic nuclei on the femtometer
scale. They imply also that the  central charge density of large nuclei is
approximately constant.  For more details, we recommend  the Nobel Lecture
(1961) by R. Hofstadter ~\cite{HofstadterNobel}.  The results summarized there
were one of the first observations of images on the femtometer scale,
corresponding to nuclear charge density distributions.  The more recent
historical overview of ref.~\cite{Glauber:2006gd} discusses applications of
multiple diffraction theory to high energy particle and nuclear physics.
Recently, with 7 and 8 TeV colliding energies of proton-proton reactions at
CERN LHC, the resolution of diffractive images in elastic proton-proton
scattering reached the sub-femtometer scales, as we demonstrate below. 

Our talk at the  Low-X 2013 conference discussed two model classes: the
Bialas-Bzdak and the Glauber-Velasco models.  Our conference contribution
follows the lines of that presentation, except for the details of on results
from the Bialas-Bzdak model, for which we direct the interested readers to
suitable references.

\section{Diffraction in quark-diquark models}

In a series of papers, Bialas and Bzdak discussed a quark-diquark model  of
elastic proton proton
scattering~\cite{Bialas:2006qf,Bzdak:2007qq,Bialas:2006kw,Bialas:2007eg}.
Recently, this Bialas-Bzdak or BB model  was tested in details on elastic
proton-proton scattering data both at the ISR and LHC
energies~\cite{Nemes:2012cp} and developed further to obtain a more realistic
description of the dip region of elastic pp scattering~\cite{CsorgO:2013kua}.

In the BB model, the differential cross-section of elastic proton-proton
scattering is given by the following formula
\begin{equation}
\frac{d\sigma}{dt}=\frac{1}{4\pi}\left|T\left(\Delta\right)\right|^2, \label{e:1}
\end{equation}
where $\Delta=|{\vec\Delta}|$ is the modulus of the transverse component of
the momentum transfer.  In the high energy limit, $s\rightarrow\infty$,
$\Delta^{2} \simeq -t$, where $t$ is the squared four-momentum transfer.  The
amplitude of the elastic scattering in momentum space, $T(\vec{\Delta})$  is
given by the Fourier-transform of the amplitude in impact parameter space,
\begin{equation}
  T(\vec{\Delta})=\int\limits^{+\infty}_{-\infty}\int\limits^{+\infty}_{-\infty}{t_{el}(\vec{b})e^{i\vec{\Delta} \cdot \vec{b}}\text{d}^2b}=
  2\pi\int\limits_0^{+\infty}{t_{el}\left(b\right)J_0\left(\Delta b\right)b {\text d}b},
  \label{e:2}
\end{equation}
where the impact parameter is denoted by ${\vec b}$ and  $b=|\vec{b}|$.  From
unitarity  conditions one obtains
\begin{equation}
  t_{el}(\vec{b})=1-\sqrt{1-\sigma(\vec{b})}.\label{e:3}
\end{equation}
The inelastic proton-proton cross-section in the impact parameter space for a
fixed impact parameter $\vec{b}$ is given by the following integral
\begin{equation}
\sigma(\vec{b})=\int\limits^{+\infty}_{-\infty}...\int\limits^{+\infty}_{-\infty}
{\text{d}^2s_q \text{d}^2s'_q \text{d}^2s_d \text{d}^2s'_d D(\vec{s_q},\vec{s_d})
D(\vec{s_q}',\vec{s_d}')
\sigma(\vec{s_q},\vec{s_d};\vec{s_q}',\vec{s_d}';\vec{b})},
\label{e:4}
\end{equation}
where the integral is taken over the two-dimensional transverse position
vectors of the quarks $\vec{s_{q}}$, $\vec{s_{q}}'$ and  diquarks
$\vec{s_{d}}$, $\vec{s_{d}}'$.

\section{Bialas - Bzdak model of elastic pp scattering}

The BB model approximates the quark-diquark distribution inside the proton
with a Gaussian
form~\cite{Bialas:2006qf,Bzdak:2007qq,Bialas:2006kw,Bialas:2007eg}
\begin{equation}
D\left(\vec{s_q},\vec{s_d}\right)=\frac{1+\lambda^2}{\pi R_{qd}^2}e^{-(s_q^2+s_d^2)/R_{qd}^2}\delta^2(\vec{s_d}+\lambda \vec{s_q}),\;\lambda=m_q / m_d,\label{e:5}
\end{equation}
and, in order to define a model that can be analytically integrated and
compared to data in a straight-forward way, the model is formulated in simple
and if possible Gaussian terms.  The BB model also supposes that protons are
scattered elastically if and only if all of their constituents are scattered
elastically
\begin{equation}
  \sigma(\vec{s_q},\vec{s_d};\vec{s_q}',\vec{s_d}';\vec{b}) =
1-\prod_{a,b \in \{q,d\}}\left[1-\sigma_{ab}(\vec{b}+\vec{s_a}'-\vec{s_b} )\right], \label{e:6} 
\end{equation}
where the inelastic differential cross-sections of the constituents are
parametrized with Gaussian distributions as well
\begin{equation}
  \sigma_{ab}\left(\vec{s}\right) = 
	A_{ab}e^{-s^2/R_{ab}^2},\;R_{ab}^2=R_a^2+R_b^2.
\label{e:7}
\end{equation}
Here $A_{ab}$ are suitably chosen normalization factors, $a,b$ stand for
$q,d$, denoting quarks and/or diquarks, while $R_q$ and $R_d$ stand for the
Gaussian radii, that characterize  in the BB model the  quark and diquark
inelastic scattering cross-sections, respectively.

\begin{figure}[H]
\includegraphics[width=0.40\textwidth]{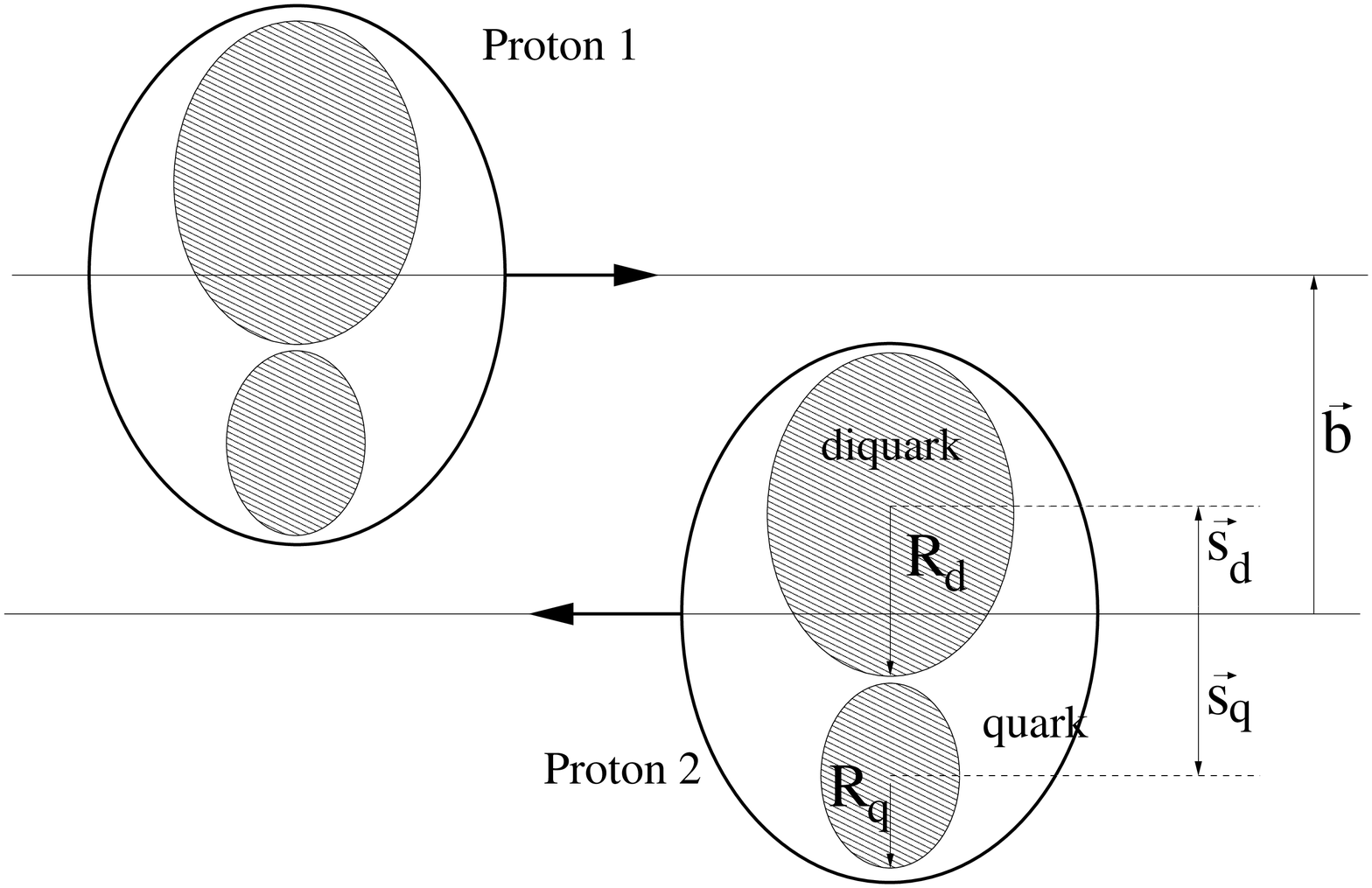}
\includegraphics[width=0.40\textwidth]{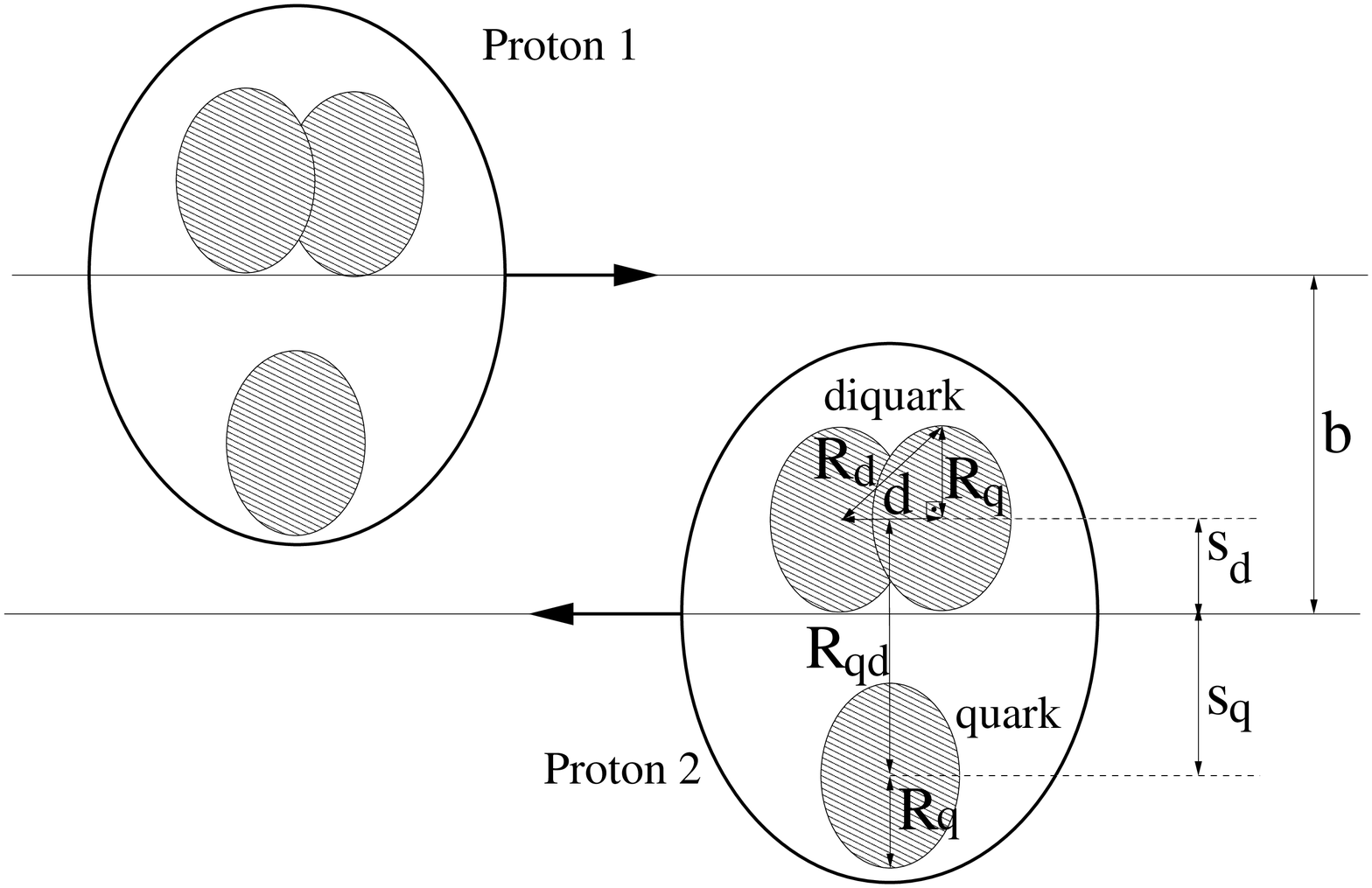}
\centering
\caption{Scheme of the scattering of two protons, when the proton is assumed
to have a quark-diquark structure. The diquark is assumed to be scattered as a
single entity (left) or as composition of two quarks (right).  This figure is
a snapshot and all the model parameters follow a Gaussian distribution. Note,
that a center of mass energy dependent Lorentz-contraction determines the
longitudinal scale parameters.}
\label{scattering sit1}
\end{figure}

This BB model comes in two different realizations, corresponding to two
different pictures of the proton: in one of the cases, the diquark is assumed
to have a structureless Gaussian distribution, while in the other case, the
diquark is assumed to scatter as a loosely bound state of two correlated
quarks. These scenarios are indicated by the $ p = (q, d) $ and the  $p = (q,
(q,q))$ labels. As noted by Bialas and Bzdak, models with three uncorrelated
quarks in the proton, labelled as $ p= (q,q,q)$ were tested before at ISR
energies and they are known to fail, with other words, we know that the quarks
are correlated inside the protons~\cite{Czyz:1969jg}.  In its original form,
the BB model has been integrated analytically, both for the $p = (q,d)$ and
the $p=(q,(q,q)) $ scenarios, {\it assuming} that the real part of forward
scattering is negligible.

\begin{figure}[H]
\includegraphics[width=0.49\textwidth]{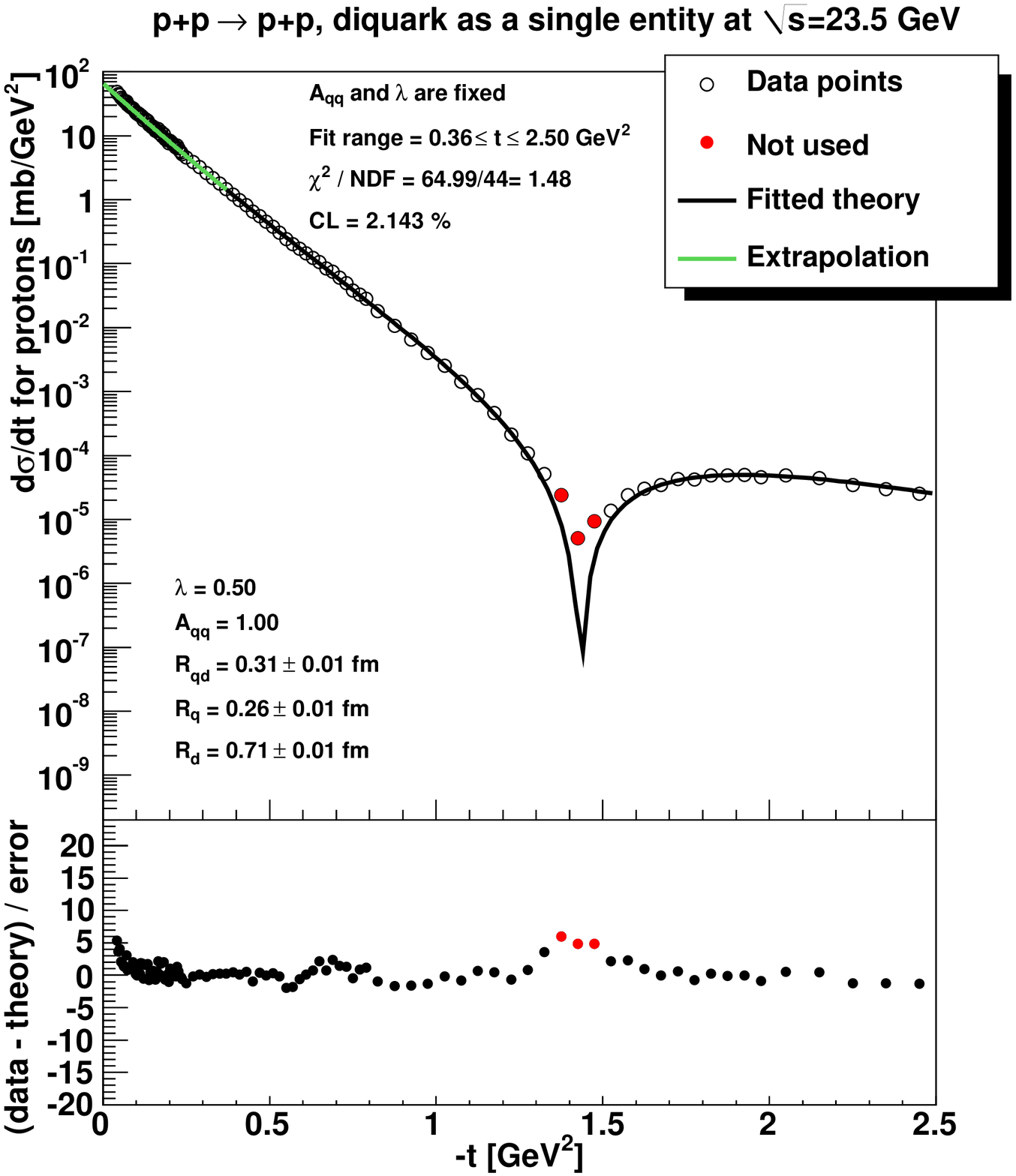}
\includegraphics[width=0.49\textwidth]{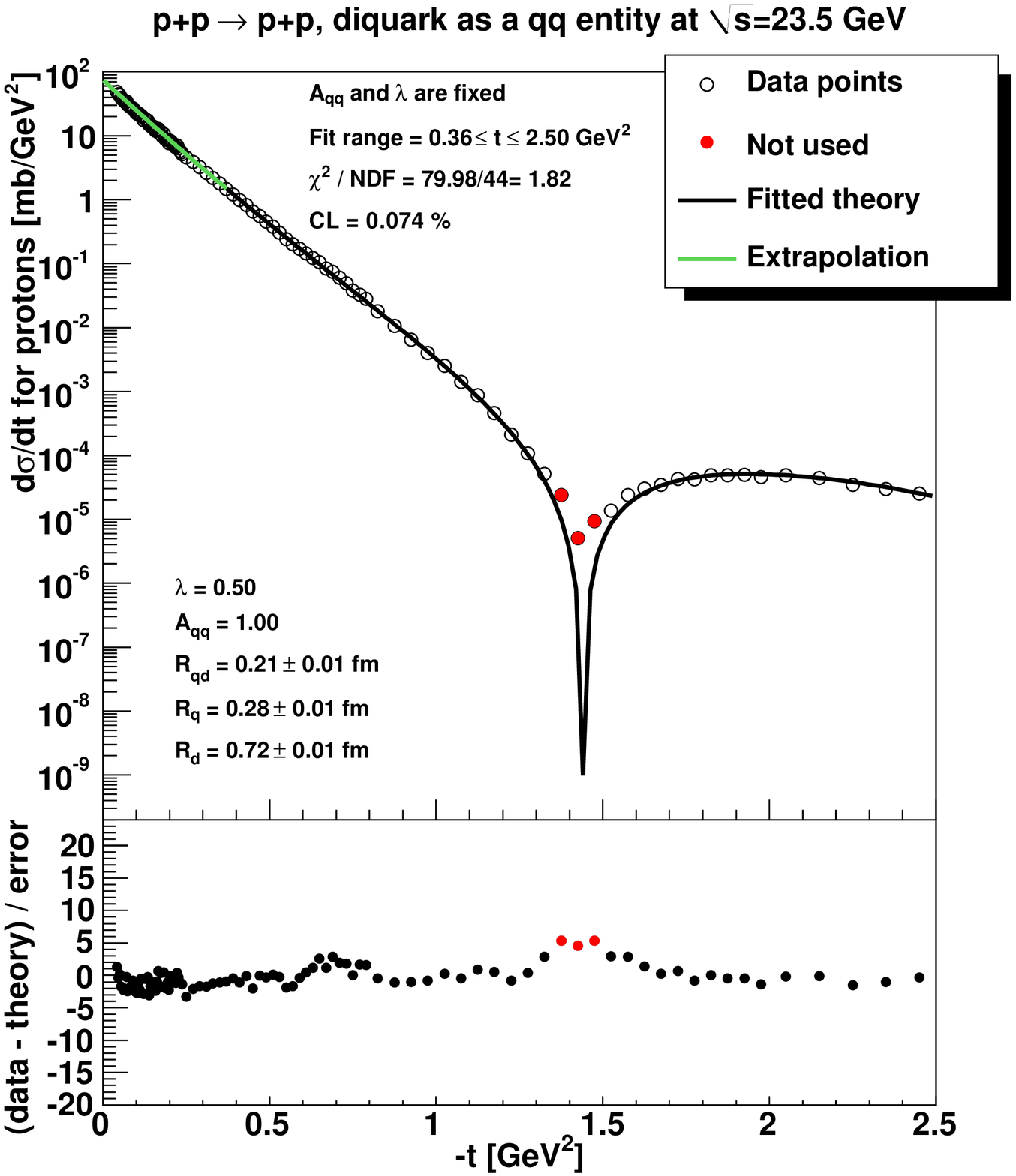}
\centering

\caption{(Color online.) Results of Minuit fits of both versions of the
Bialas-Bzak model at ISR energies. Left panel indicates the scenario $p =
(q,d)$, where the diquark is assumed to scatter as a single entity while the
right panel indicates the scenario $p = (q,(q,q))$, where the diquark inside
the proton is considered to be a scattering object consisting of two quarks.}
\label{Figure:2}
\end{figure}

\begin{figure}[H] 
\includegraphics[width=0.49\textwidth]{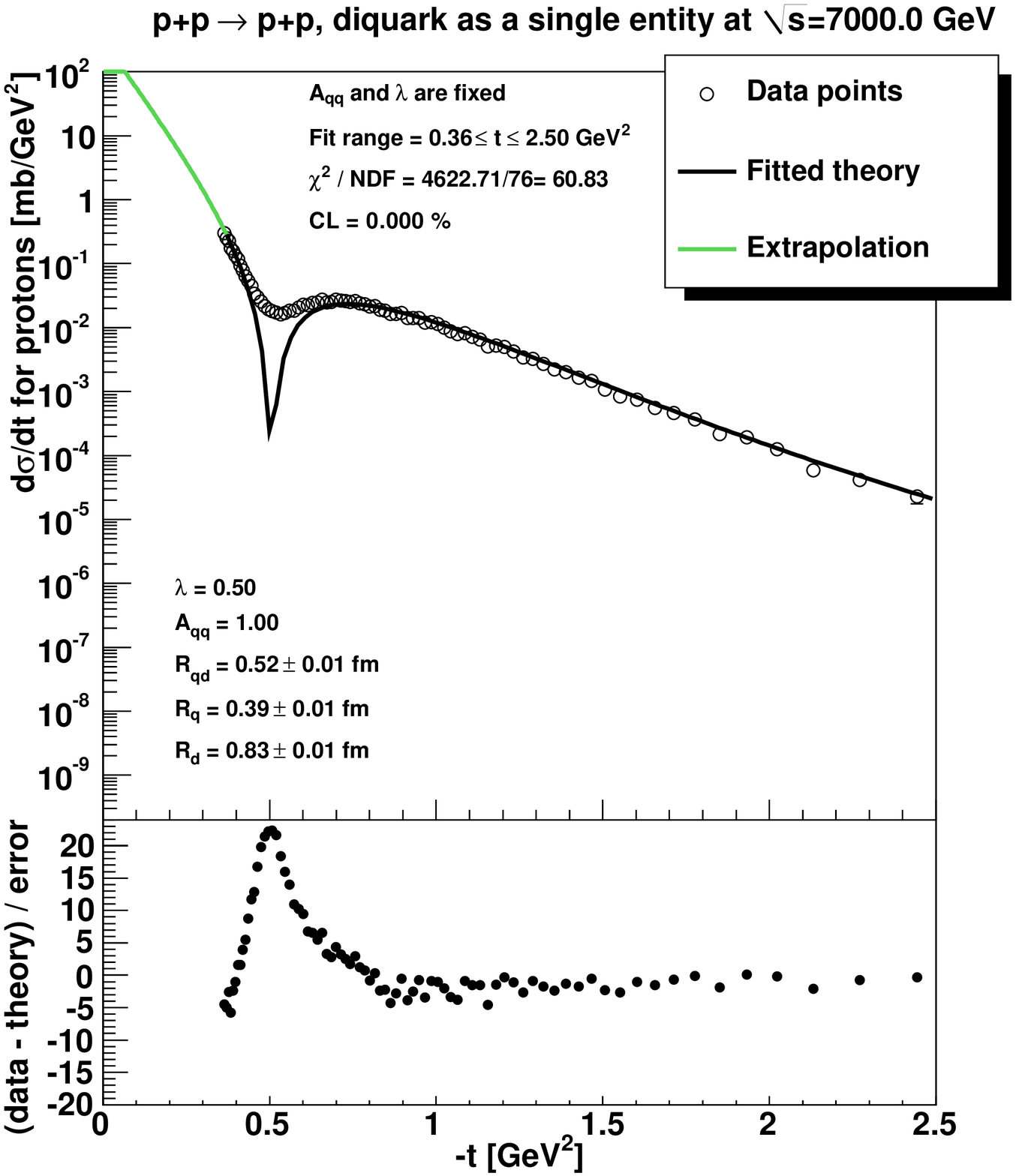} 
\includegraphics[width=0.49\textwidth]{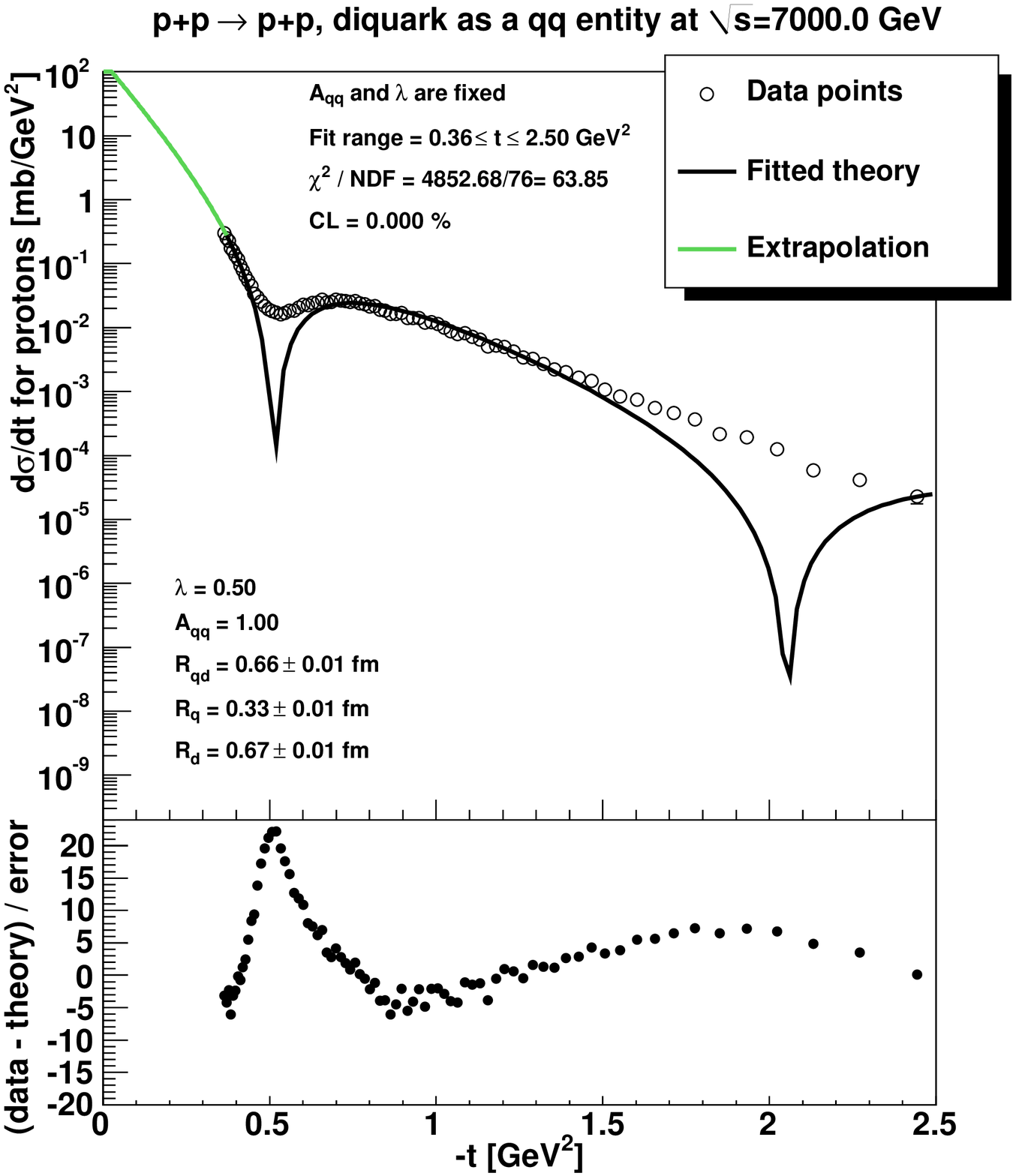} 

\centering 
\caption{(Color online.) The result of the fit of the original
version of the Bialas-Bzdak model at $7$ TeV LHC energies in two different
scenarions: left panel stands for the $p=(q,d)$ scenario, when the diquark is
assumed to scatter as a single entity, while the right panel stands for the
$p=(q,(q,q))$ case, when the internal structure of the diquark is resolved as
a correlated system of two quarks. } 
\label{Figure:3} 
\end{figure}

The two panels of Figure ~\ref{Figure:2} indicate CERN Minuit fit results of
the BB model to differential cross-section data on elastic proton-proton
scattering at the ISR energy of $\sqrt{s}= 23.5$ GeV.  Left plots correspond
to the scenario $p = (q, d)$ while the right panel stands for the scenario $p
= (q, (q, q))$.  The top parts show the data points and the result of the best
fit, while the lower parts indicate the relative deviation of the model from
data in units of measured error bars.  As the original BB model is singular
around the dip, 3 data points were left out from the optimalization, that are
located closest to the diffractive minimum, and indicated with filled (red)
circles in Figure~\ref{Figure:2}.  The fit range was restricted to $0.36 \le
- t \le 2.5$ GeV$^2$, so that a fair comparison could subsequently be made
with the first TOTEM results  on proton-proton elastic scattering at LHC
energy of 7  TeV of ref. ~\cite{Antchev:2011zz}. The best fits are shown with
a solid (black) line in the fitted range, while their extrapolation to lower
$t$ values are also shown, with dashed (green) lines. The confidence levels,
after fixing the values of $\lambda$ and $A_{qq}$ to 0.5 and 1, respectively,
come very close to 0.1\%, indicating that the fit quality is similar,
statistically acceptable in both scenarios. Similar fit qualities were
reported at each ISR energies of 30.7 GeV, 52.8 GeV and 62.5 GeV, see ref.
~\cite{Nemes:2012cp} for details. Figure ~\ref{Figure:3} shows the comparison
of the BB model to TOTEM data on elastic pp scattering at 7 TeV LHC energies,
indicating a qualitative change, as compared to the fit results at ISR
energies: the quality of this fit is statistically not acceptable, CL is
significantly below 0.1\%, and the fit deviates from the data in particular in
the dip region. The bottom parts indicate, that the shape of the differential
cross-section in the dip region, around the first diffractive minimum is not
reproduced correctly by the original BB model at 7 TeV LHC energies, and, as
also can be seen on this Figure ~\ref{Figure:3}, this shortcoming cannot be
fixed by leaving out a few data points around the diffractive minimum from the
optimalization procedure. The details of these BB fits are described in ref.
~\cite{Nemes:2012cp}. 

 Recently, two of us generalized the Bialas-Bzdak model by adding a small real
part to the forward scattering amplitude, to investigate, if the description
of the dip region can be improved can be made statistically acceptable in this
way. The results of this scenario are described in detail in
ref.~\cite{CsorgO:2013kua}. A {\it small} real part was added to the forward
scattering amplitude by using an analogy of with the Glauber-Velasco model,
and assuming that even if all the parton level scatterings are elastic, the
proton-proton scattering can, with a small probability, become inelastic. In
this manner, a parton level $\rho$ parameter was introduced.  The results,
detailed in ref.~\cite{CsorgO:2013kua}, indicate that a small real part indeed
improves the agreement of the BB model with data in the dip region, and the
fits become statistically acceptable in the whole $t$ region, including all
the data points from dip region, {\it if} the energy of the collisions is
limited to the ISR energy range of $\sqrt{s} = 23.5 $ - $ 62.5 $ GeV. At the
LHC energy of $\sqrt{s} = 7$  TeV, the generalized Bialas-Bzdak or the
$\alpha$BB model resulted in an improvement, that reduced the disagreement
between the BB model and the data substantially and filled the dip region
rather dramatically. However, even this improvement did not result in a
statistically acceptable fit quality to the differential cross-section of
elastic proton-proton collisions at this LHC energy, although good quality
fits were obtained if the fit range was limited to the dip region. As a
consequence, we kept on searching for a model that is able to describe elastic
pp scattering data at LHC energies, and investigated the performance of the
Glauber-Velasco model~\cite{Glauber:1984su}. Before reporting the results, let
us summarize what we have learned till now from the detailed fits using the
original Bialas-Bzdak or BB model, and its generalized version when a small
real part is added to its forward scattering amplitude,
see refs.~\cite{Nemes:2012cp,CsorgO:2013kua} for further details. 

\section{What have we learnt so far ?} 

The original version of the Bialas-Bzdak model gave a statistically acceptable
description of elastic pp scattering data at ISR energies, if the data points
close to the diffractive minimum were left out from the fit. If these data
points were included and also a small real part was added to the model, as
detailed in ref.~\cite{CsorgO:2013kua}, the fits at the ISR energies from
$\sqrt{s} =  $ 23.5 GeV to 62.5 GeV become statistically acceptable, good
quality fits, in the fit range of 0.36 $\le -t \le $ 2.5 GeV$^2$. Two model
parameters could be fixed at all energies ($A_{qq} = 1$ and $\lambda = 1$)
while maintaining the statistically acceptable fit quality. The parameter
$\alpha$, that was introduced as a parton level ratio of the real to imaginary
part of the forward scattering amplitude, remained indeed in the region of
very small values, $\alpha = 0.01 \pm 0.01$ except at 52.8 GeV, where $\alpha
= 0.02 \pm 0.01$ value was found. Although these $\alpha$ parameters are
within errors consistent with zero, a small but non-vanishing value provided
qualitatively better fits in the dip region, as detailed in
ref.~\cite{CsorgO:2013kua}. The best fit parameters, that described the quark
structure of the protons geometrically, took also rather interesting values.
For example, the  quark radius $R_q$ within 2 standard deviations was
consistent with an energy independent value of $R_q = 0.27 \pm 0.01$ fm. The
diquark size indicated a nearly constant value, varying between $R_d = 0.71
\pm 0.01$ to $0.77 \pm 0.01$ fm, slighlty increasing with increasing
$\sqrt{s}$. Although the fit to the TOTEM data 7 TeV were not statistically
acceptable, the best parameter values for the quark and diquark radii were in
the same range, except a slight decrease of the diquark size in the $p =
(q,(q,q))$ model at 7 TeV. We observed that the biggest variation, when the
energy is increased to 7 TeV, is observable in the scale that measures the
typical quark-diquark distance, $R_{qd}$. This value was in the range of
$R_{qd} = 0.23\pm 0.01$ fm at ISR energies in the $p = (q, (q,q)) $ model,
while it increased to the value of $0.73 \pm 0.01 $ fm at 7 TeV. Similar trend
of increasing quark-diquark separation is seen in the $p = (q,d)$ scenario.
Graphically, the evolution of the proton elastic scattering structure is
illustrated on Figure~\ref{Figure:3}, where the best fit parameters are also
indicated on the sub-plots, generated for the case of the $p=(q,(q,q))$
scenario.  The same qualitative behaviour of increasing quark-diquark distance
is observed also in the $p = (q, d)$ picture, see ref.~\cite{CsorgO:2013kua}
for further details. 

\begin{figure}[H] 
\centering 
\includegraphics[trim = 6mm 12mm 2mm 4mm, clip, width=0.4\linewidth]{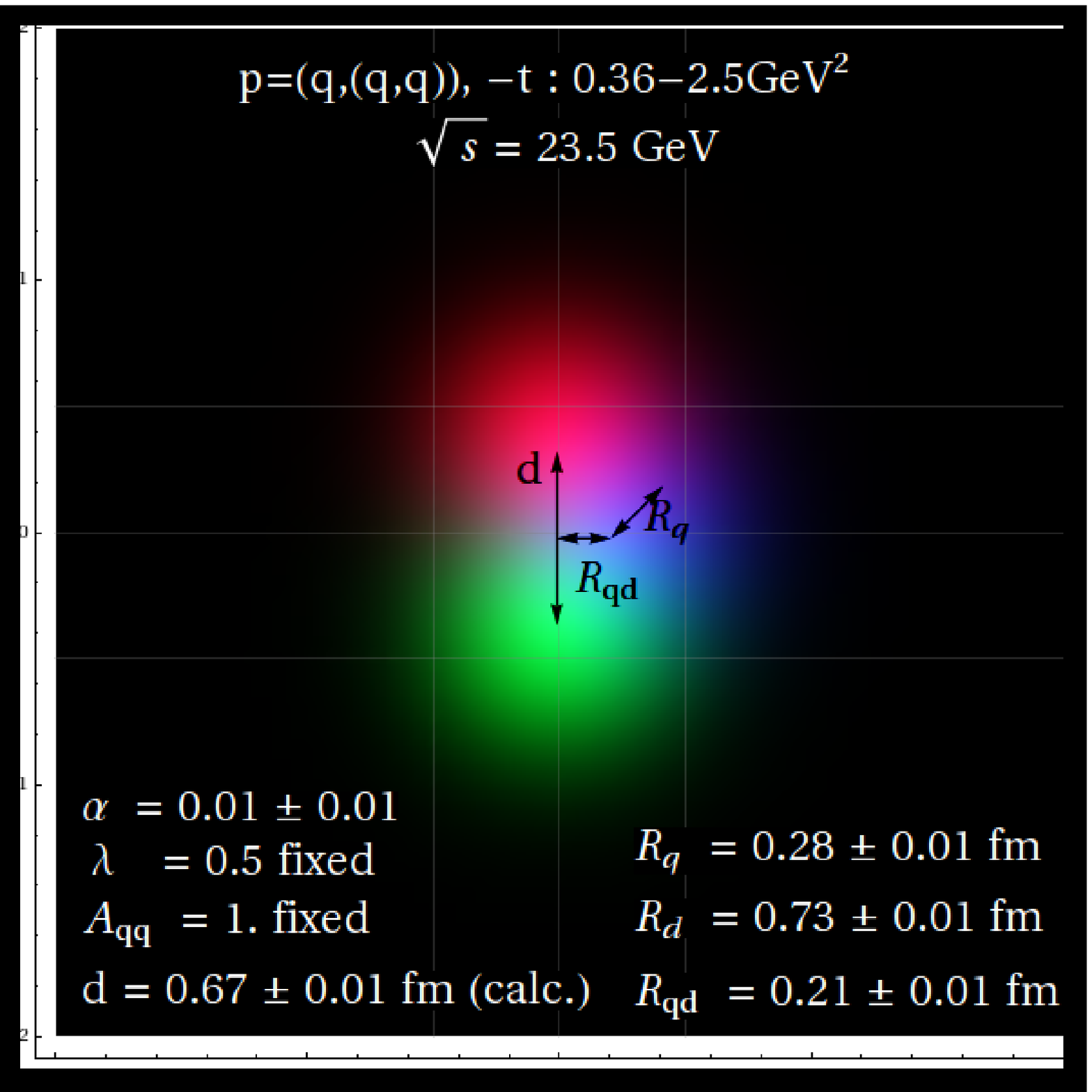} 
\includegraphics[trim = 6mm 12mm 2mm 4mm, clip, width=0.4\linewidth]{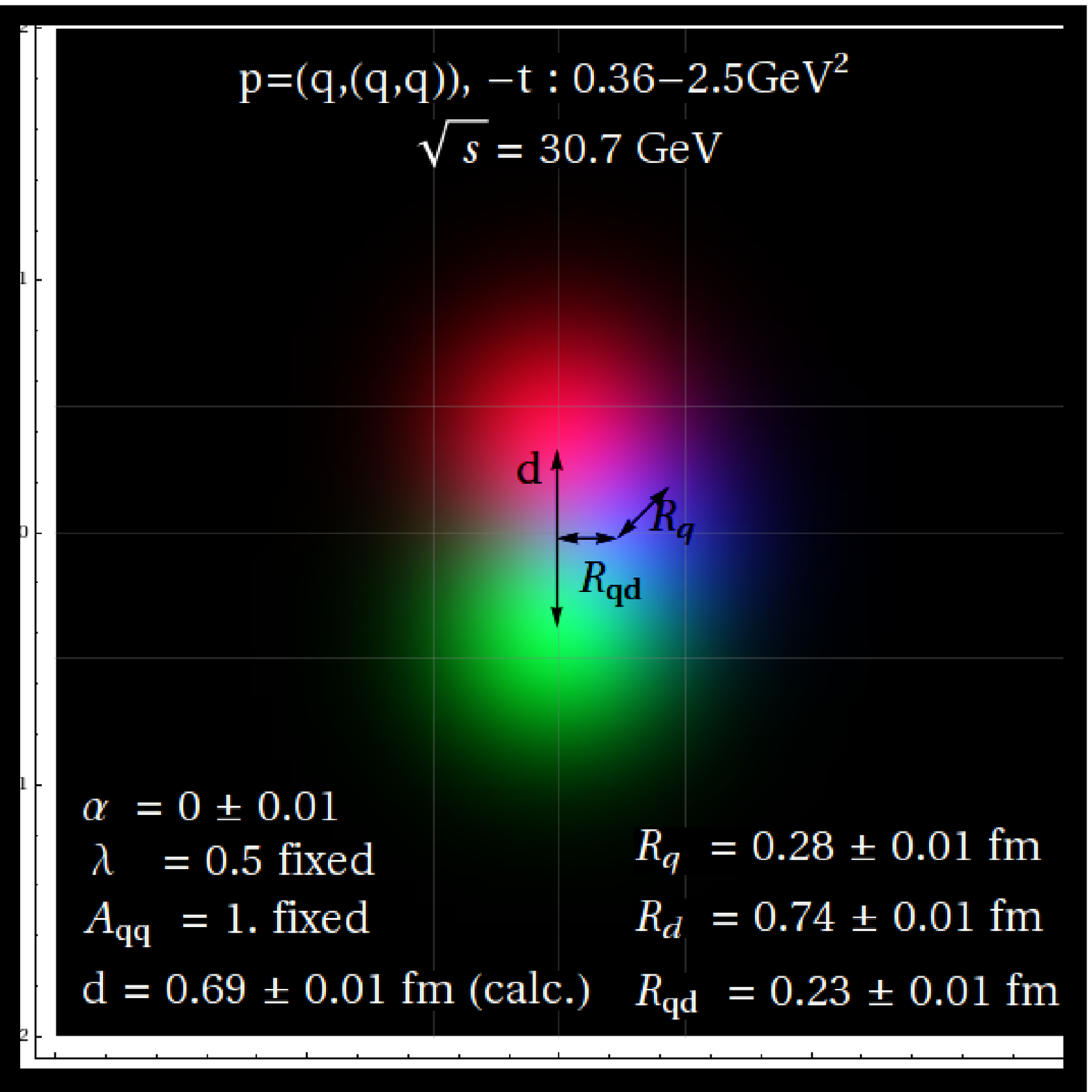}\\
\includegraphics[trim = 6mm 12mm 2mm 4mm, clip, width=0.4\linewidth]{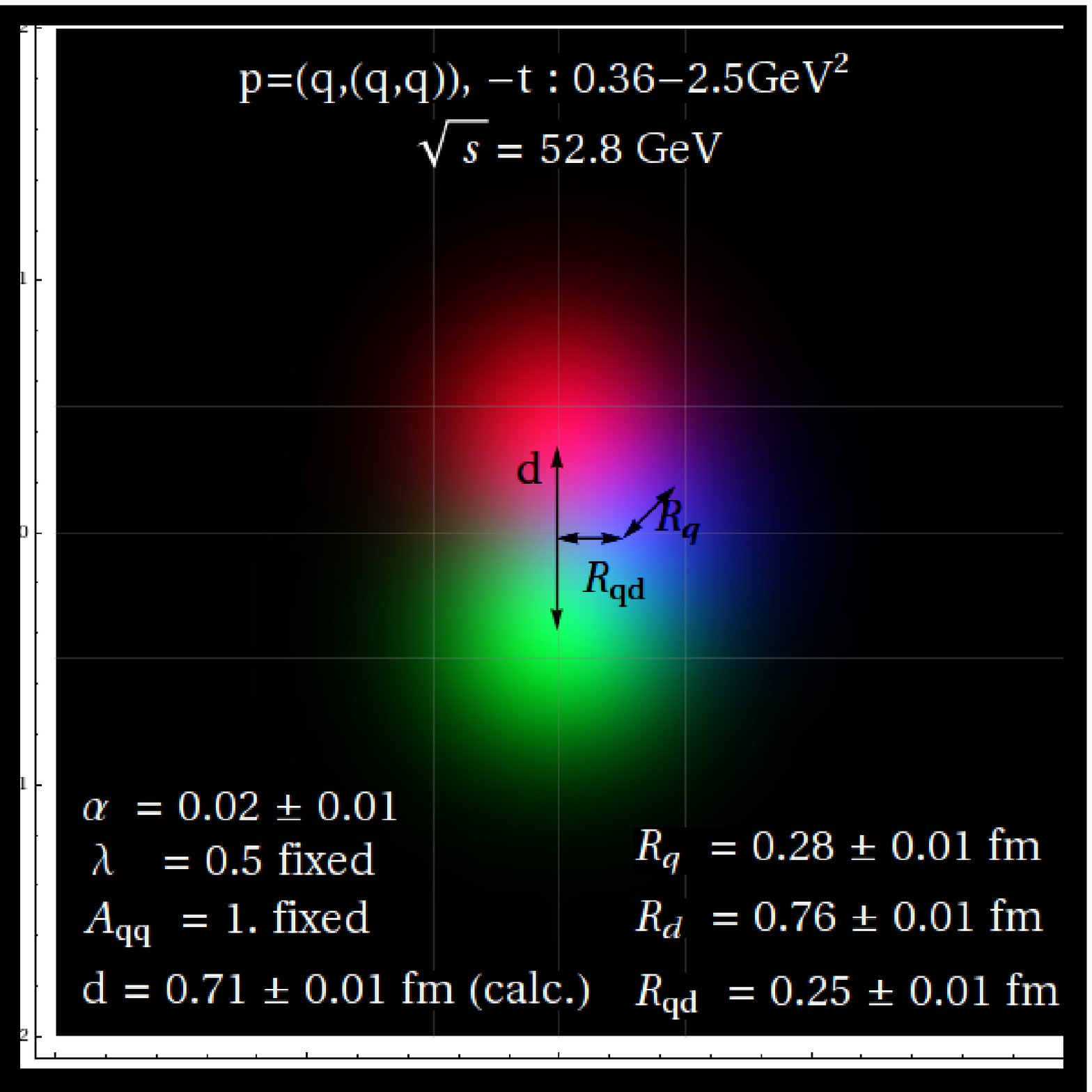} 
\includegraphics[trim = 6mm 12mm 2mm 4mm, clip, width=0.4\linewidth]{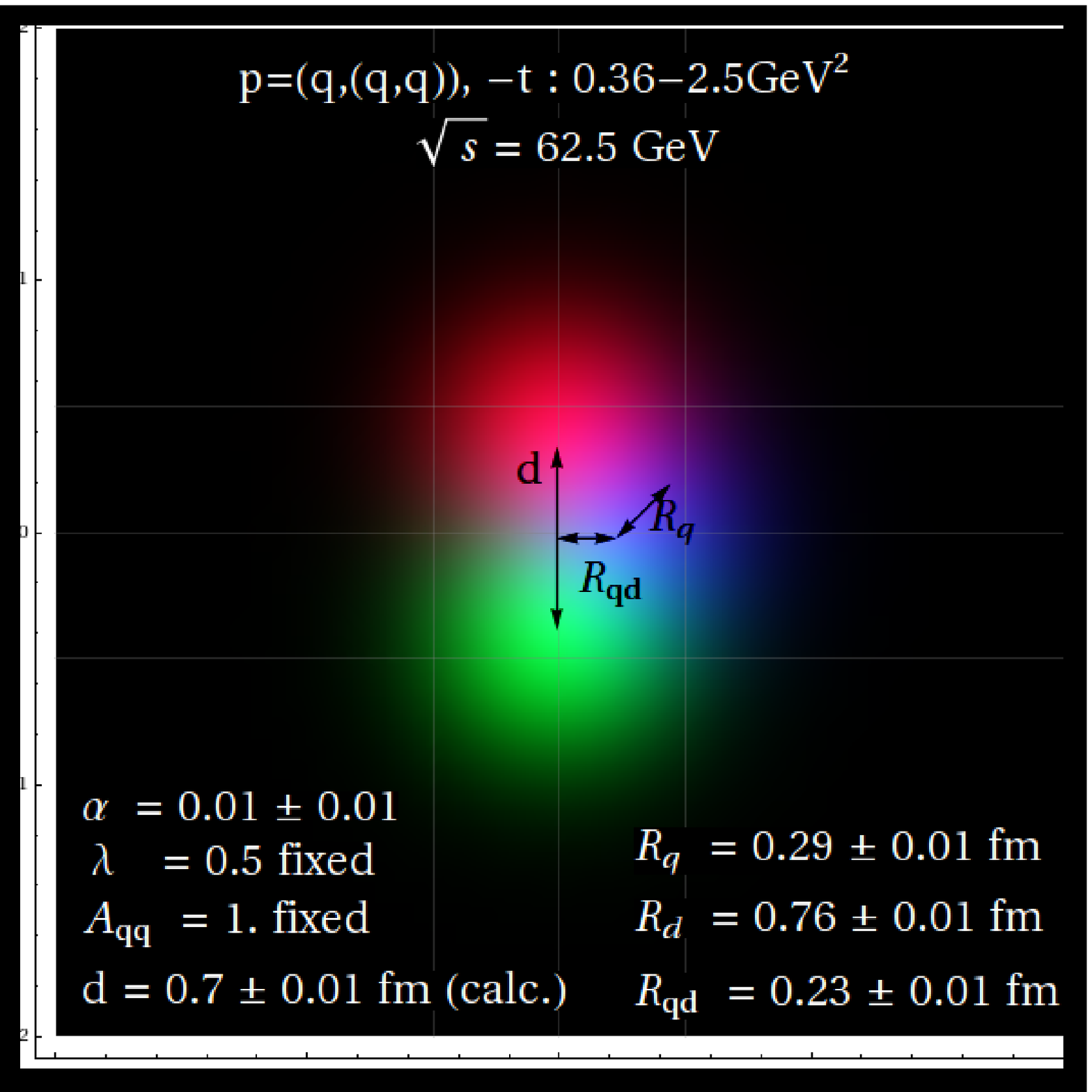}\\
\includegraphics[trim = 6mm 12mm 2mm 4mm, clip, width=0.4\linewidth]{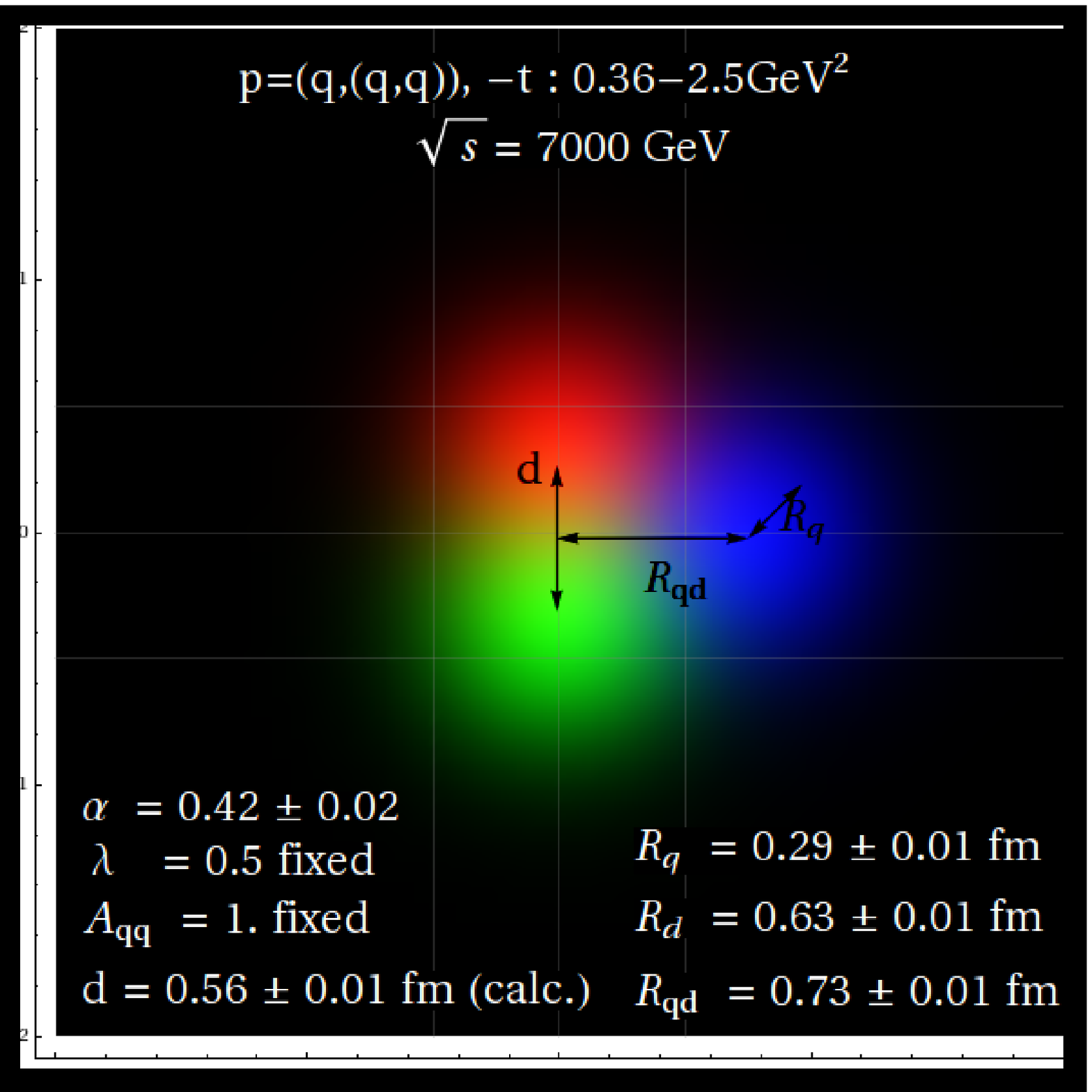} 
\caption{Visualization of the fit results of Bialas-Bzdak models, extended to
a small real part, for the case of $p = (q, (q, q)) $, when the diquark is
assumed to be resolvable as a weekly bound state of two quarks. The main
effect of increasing $\sqrt{s}$ is apparently the increasing value of
$R_{qd}$, the typical quark-diquark distance.} 
\label{Figure:4} 
\end{figure} 

As discussed both in refs.~\cite{Nemes:2012cp} and ~\cite{CsorgO:2013kua}, the
$p = (q,d)$ and the $p = (q, (q,q))$ models provide similar quality of data
description both at ISR energies (where they are both statistically
acceptable) and at 7 TeV LHC energy (where both fail to describe TOTEM data in
a statistically acceptable manner). Nevertheless we compare the best fit
values of the different energies, to try to get a qualitative insight
assuming, that the missing element of the model will not modify drastically
the best fit parameters at LHC energies. Given that the $p = (q,d)$ and the $p
= (q,(q,q))$ Bialas-Bzdak models  correspond to two different assumptions
about the internal structure of the protons, it was a kind of surprize for us,
that the measured total pp cross-section $\sigma_{tot}$ was phenomenologically
related to the parameters of the BB model in a model-independent way, i.e. the
following relation is approximately valid for both scenarios: 

\begin{equation} 
\sigma_{tot} \approx 
	2 \pi R_{eff}^2 \,\, = \,\, 2 \pi (R_q^2 + R_d^2 + R_{qd}^2). 
\end{equation} 

This approximation was found to be valid within a relative error of about 9 \%
at ISR energies, while at the LHC energies it yields only an ball-park value,
order of magnitude estimation ($\frac{\sigma_{tot}}{2 \pi R_{eff}^2} = 1.42$).
We also have observed an interesting scaling property of the differential and
the total proton-proton elastic scattering cross-section, namely the product
of the total cross-section times the $t$ of the dip is within errors a
constant: \begin{equation} t_{dip}\sigma_{tot} \approx C \end{equation} where
$C = 54.8 \pm 0.7$  mb GeV$^2$ from a fit. We find that this relation is valid
within 5 \% relative error at each ISR and also at 7 TeV LHC energies. A
similar relation holds for a light scattering from a black disc, however, with
a significantly different constant value, $C_{blackdisc} \approx 35.9 $ mb
GeV$^2$. 
\begin{figure}[H] 
\centering
\includegraphics[width=0.48\textwidth]{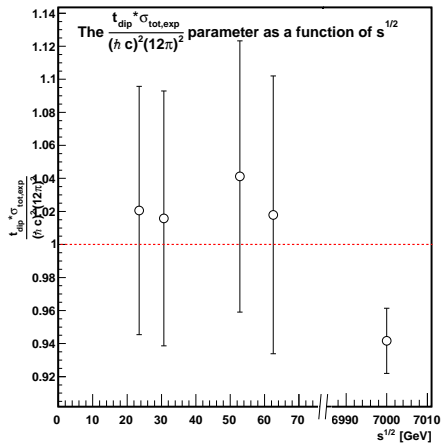} 
\caption{ The $\frac{|t_{dip}| \cdot \sigma_{tot,exp}}{C}$ ratio, 
directly obtained from experimental data.
The dashed line indicates $1$, which value within errors is consistent with
all the data from $\sqrt{s} = 23.5 $ GeV to 7 TeV. } 
\label{Figure:5}
\end{figure} 

Given that there are theoretically well established formulas for
the description of the rise of the total pp scattering cross-section with
increasing energies, the above formula can be well used to predict the
position of the (first) minimum or the dip in the differential cross-section
of pp collisions and also can be extrapolated, or, predicted for pA and AB
collisions ~\cite{inprep}. Given that we could not find a statistically
acceptable quality fit with the Bialas-Bzdak model to 7 TeV TOTEM data on
elastic pp scattering at LHC, neither in the original form, nor when a small
real part is added to the forward scattering amplitude of this model, we
started to look for alternative interpretations and derivations of
$d\sigma/dt$. One possibility is to allow for not only small values of the
real part of the forward scattering amplitude, but still keep the basic
structure of the Bialas-Bzdak model. The studies in this direction will be
reported elsewhere. In the next section we report about the other natural
direction, that we investigated in detail. In particular, when we added a
small real part to the forward scattering amplitude to the Bialas-Bzdak model
in ref.~\cite{CsorgO:2013kua}, we were introducing a parton level $\rho$
parameter inspired by the Glauber-Velasco model of
refs.~\cite{Glauber:1984su,Glauber:1987sf}. In the next section, we summarize
this model and report about its first comparisions to TOTEM data. 

\section{Overview of the model of Glauber and Velasco} 

In this section, we follow the lines of the presentation of the
Glauber-Velasco model, as described in
refs.~\cite{Glauber:1984su,Glauber:1987sf}. The Glauber  diffractive multiple
scattering theory is utilized to describe elastic collisions of two nucleons,
which are pictured as clusters of partons. The parton distributions are
assumed to have form factors given by the experimentally measured electric
charge form factors. Differential cross sections calculated in this way showed
good agreement with the experimentally measured ones over a broad range of
$pp$ and $p\bar{p}$ energies, when the parton-parton scattering amplitude is
given a suitable parametrization~\cite{Glauber:1984su,Glauber:1987sf}. The
range of the parton-parton interaction derived from these data is found to
increase steadily with energy. The absorption processes that take place are
localized in the overall nucleon-nucleon interaction by calculating the shadow
profile function. The emerging picture corresponded to an opaque region of
interaction that grows in radius with increasing energy. The surface region of
the interaction seems however to maintain a remarkably fixed shape as the
radius grows. In the multiple diffraction theory of Glauber and Velasco, the
elastic scattering amplitude for diffractive collisions can be written as an
impact parameter integral 

\begin{equation} 
F(t)=i\int_0^{\infty} J_0\left(b\sqrt{-t}\right)\left\{1-\exp\left[ -\Omega\left(b\right)\right]\right\}bdb.
\end{equation} 

Any particular model is characterized by the opacity function $\Omega(b)$,
which in general may be a  complex valued function. If we picture the two
colliding nucleons as clusters of partons that scatter one another with the
averaged scattering amplitude $f(t)$, then the opacity function can be written
in the form of an integral over momentum transfers $q$, 

\begin{equation} 
\Omega\left(b\right)=\frac{\kappa}{4\pi}
\left(1-i\alpha\right)\int_0^{\infty}
{J_0\left(q\,b\right)G_{p,E}^2\left(-t\right)}\frac{f(t)}{f(0)}q dq,
\end{equation} 

where $q = \sqrt{-t}$. The constants $\kappa$ and $\alpha$ in this expression
are real-valued and need to be determined empirically. The function
$G_{p,E}(t)$ is the form factor for the parton density in the proton.
Following ref.~\cite{Glauber:1987sf}, we shall assume it to be the same as the
observed electric form factor for the proton. One choice of parametrization we
have investigated is 

\begin{equation} 
\frac{f(t)}{f(0)} = \frac{e^{i\left(b_1\left|t\right|+b_2\,t^2\right)}}
{\sqrt{1+a\left|t\right|}}.\\ 
\end{equation} 

The BSWW form factor, corresponding to the distribution of electric charge in the proton, is described with a four-pole parametrization~\cite{Borkowski_1975} 

\begin{equation} 
G_{p,E}\left(q^2\right)=\sum_{i=1}^{n}\frac{a_i^{E} \left(m_{i}^{E}\right)^2} {\left(m_{i}^{E}\right)^2 + q^2}\,,\, \sum_{i=1}^{n}a_i^E=1\,,\, G_{p.E}(0)=1.\\ \label{BSWW_four_pol_parametrization} 
\end{equation} 

The differential cross-section for elastic pp collisions is evaluated as 
\begin{equation} 
\frac{d\sigma_{el}}{d\left|t\right|}=\pi\left|F(t)\right|^2. 
\end{equation} 

The parameters of the BSWW form factor are given by the following table: 
\begin{table}[!h] \centering 
\begin{tabular}{|r|r|} 
\hline $a_i^E$ & $(m_i^E)^2$ (fm$^{-2}$) \\ 
\hline 0.219 	& 3.53 		\\ 
\hline 1.371 	& 15.02		\\ 
\hline -0.634	& 44.08		\\ 
\hline 0.044 	& 154.20	\\ 
\hline 
\end{tabular} 
\caption{Best fit parameters~\cite{Borkowski_1975} of the four-pole fit of
Eq.~(\ref{BSWW_four_pol_parametrization}).} 
\end{table} 

\begin{figure}[H] 
\centering 
\includegraphics[width=0.49\linewidth]{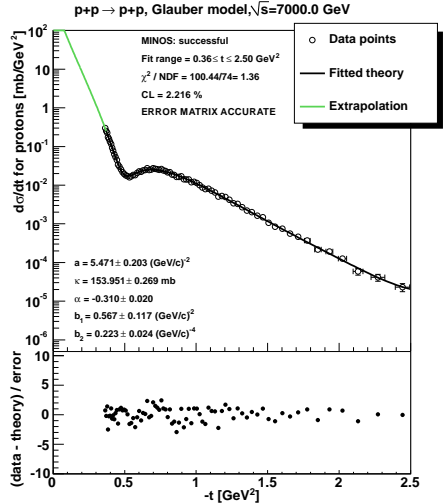} 
\caption{Glauber-Velasco model fit to the differential cross-sectio  of  proton-proton elastic scattering data at $7$ TeV, in the range of $0.36 \le -t \le 2.5 $ GeV$^2$.
} 
\label{Figure:6} 
\end{figure} 

Figure~\ref{Figure:6} indicates, that the Glauber-Velasco model is able to
describe successfully the differential scattering cross-section of  elastic pp
collisions at the 7 TeV LHC energies: the fit quality is statistically
acceptable, with CL  $> 0.1$ \%. We have tested the model at the ISR energy
range of 23.5 GeV - 62.5 GeV too, where the similarly good quality fits were
found. The detailed results will be reported in a manuscript that is currently
under preparation. 

\section{Summary} 
In summary, we have analized elastic
proton-proton scattering data from the 23.5 GeV ISR energies to 7 TeV LHC
energies, using various forms of the Bialas- Bzdak model. We found that the
scenario when the proton is considered to be a quark-diquark state provides a
fit quality that is similar to the case when the diquark is resolved as a
correlated quark-quark system within the framework of the same model. Adding a
small real part to the forward scattering amplitude of the original
Bialas-Bzdak model provides a statistically acceptable description of elastic
pp scattering data at the ISR energies, however, even this generalized
Bialas-Bzdak model fails to describe TOTEM data on elastic pp scattering at 7
TeV. Given that the generalization of the Bialas-Bzdak model followed the lines
of the Glauber-Velasco model, we tested also the performance of the
Glauber-Velasco model in its original form, and found that it was describing
elastic proton-proton scattering both at ISR and at LHC energies when the fit
range was restricted to $ 0.36 \le -t \le 2.5$ GeV$^2$. 

\section{Acknowledgments} 
T. Cs. would like to thank for professor Glauber for
his kind hospitality at Harvard University as well as for his inspiring and
fruitful visits to Hungary and CERN, that made the completion of these results
possible. He also would like to express his gratitude to the organizers of the
Low-X 2013 conference for an invitation and for creating an inspiring  and
useful meeting. This research was supported by the Hungarian OTKA grant NK
101428 and by a Ch. Simonyi fund, as well as by the Hungarian Academy of
Sciences, by a HAESF Senior Leaders and Scholars fellowship and by the US DOE.

\end{document}